\documentclass[letterpaper, 10 pt, conference]{ieeeconf}  
\usepackage{tikz}
\usepackage{lipsum}
\newcommand\copyrighttext{%
  \footnotesize Copyright~\copyright~2021 IEEE. Personal use of this material is permitted. Permission from IEEE must be obtained for all other uses, in any current or future media, including reprinting/republishing this material for advertising or promotional purposes, creating new collective works, for resale or redistribution to servers or lists, or reuse of any copyrighted component of this work in other works.}
\newcommand\copyrightnotice{%
\begin{tikzpicture}[remember picture,overlay]
\node[anchor=south,yshift=10pt] at (current page.south) {\fbox{\parbox{\dimexpr\textwidth-\fboxsep-\fboxrule\relax}{\copyrighttext}}};
\end{tikzpicture}%
}
\IEEEoverridecommandlockouts                              
\usepackage[utf8]{inputenc}
\usepackage[T1]{fontenc}

\usepackage{graphicx}

\title{\LARGE \bf
Removing Noise from Extracellular Neural Recordings Using Fully Convolutional Denoising Autoencoders
}

\author{Christodoulos Kechris$^{1\,*}$, Alexandros Delitzas$^{1\,*}$, Vasileios Matsoukas$^{1\,*}$, Panagiotis C. Petrantonakis$^{2}$
\thanks{*These authors contributed equally}
\thanks{$^{1}$Department of Electrical and Computer Engineering, Aristotle University of Thessaloniki, GR 54124, Thessaloniki, Greece}%
\thanks{$^{2}$Information Technologies Institute, Centre for Research and Technology – Hellas (CERTH), GR 57001, Thessaloniki, Greece. (correspondence e-mail: ppetrant@iti.gr)}%
}

\begin{document}

\maketitle
\thispagestyle{empty}
\pagestyle{empty}

\begin{abstract}
Extracellular recordings are severely contaminated by a considerable amount of noise sources, rendering the denoising process an extremely challenging task that should be tackled for efficient spike sorting. To this end, we propose an end-to-end deep learning approach to the problem, utilizing a Fully Convolutional Denoising Autoencoder, which learns to produce a clean neuronal activity signal from a noisy multichannel input. The experimental results on simulated data show that our proposed method can improve significantly the quality of noise-corrupted neural signals, outperforming widely-used wavelet denoising techniques.
\end{abstract}

\copyrightnotice
\section{INTRODUCTION}

Monitoring the activity of single or multiple neurons, by placing electrodes and registering extracellular recordings, is a key factor in understanding brain's mechanisms \cite{Markovic2012}. The desired goal when analyzing extracellular recordings is to identify which neurons activate by firing spikes, a technique known as Spike Sorting (SS) \cite{Quiroga2007}. The SS process traditionally involves the following steps \cite{Quiroga2015}. First, the recorded raw data are filtered since noise contamination, attributed mainly to nearby neural activity but also to hardware \cite{Markovic2012}, can negatively influence the SS performance \cite{Lewicki1998}. Second, a spike detection procedure is applied by manually or automatically setting an amplitude threshold on the filtered data. Third, a feature extraction step aims to separate the spike shapes. Finally, a clustering algorithm is utilized to group spikes with similar features into clusters, distinguishing the spike's source neuron.

For the filtering step, a simple bandpass filter in the range between 300 and 3000 Hz is commonly employed. The lower cutoff frequency removes the slow components related to Local Field Potential (LFP), while the upper prevents the visual distortion of spikes from high frequency noise. However, the use of standard bandpass filtering inserts distortions to the waveform shapes~\cite{Wiltschko2008}. To overcome this limitation, wavelet denoising techniques have been widely adopted in neural signal processing. In these techniques, the signal is represented in the time-scale domain, thereby allowing a fine analysis of the signal’s frequency components without loss of temporal information, in contrast to the traditional Fourier transform~\cite{Baldazzi2020}. Two of the most prominent wavelet denoising methods for neural signals are the Discrete Wavelet Transform (DWT) and the Stationary Wavelet Transform (SWT) \cite{Brychta2007,Baldazzi2020}. 
 
While traditional methods have performed fair enough, Deep Learning (DL) approaches have attracted attention because of their capacity to comprehend complex data representations. Biological systems have been intensively addressed by sophisticated DL approaches with impressive results including physiological signal denoising. In \cite{Koch2020}, a deep encoder-decoder algorithm is used for denoising 2D images of neuronal physiology, drastically improving the Signal-to-Noise Ratio (SNR) distribution. Convolutional Autoencoders have also showed promising potential in electoencephalogram (EEG) filtering, surpassing the traditional bandpass approach \cite{leite2018deep}. Moreover, in \cite{Arsene2020}, Convolutional Neural Networks (CNN) and Long Short-Term Memory (LSTM) networks outperformed traditional wavelet methods in electrocardiogram (ECG) denoising. In \cite{chiang2019noise} the authors experimentally show the superiority of the Fully Convolutional Denoising Autoencoder (FC-DAE) over the Dense and Convolutional Denoising Autoencoders in ECG noise reduction.

In our work, we propose a Fully Convolutional Denoising Autoencoder (FC-DAE) to eliminate noise distortion in extracellular neural activity recordings and retrieve a signal as close to ground truth as possible. We base our experiments on simulated datasets, evaluate our approach with respect to SNR and Root Mean Square Error (RMSE) metrics and compare with Wavelet Transform methods. 

All the datasets and the code discussed in this paper are publicly available at Github\footnote{https://github.com/AlexDelitzas/fcdae-neural-signal-denoising}. 

\section{Materials and Methods}

\subsection{Proposed Autoencoder Network}

\begin{figure*}[thpb]
  \centering
  \framebox{\parbox{6.5in}{
  \centering
  \includegraphics[scale=0.9]{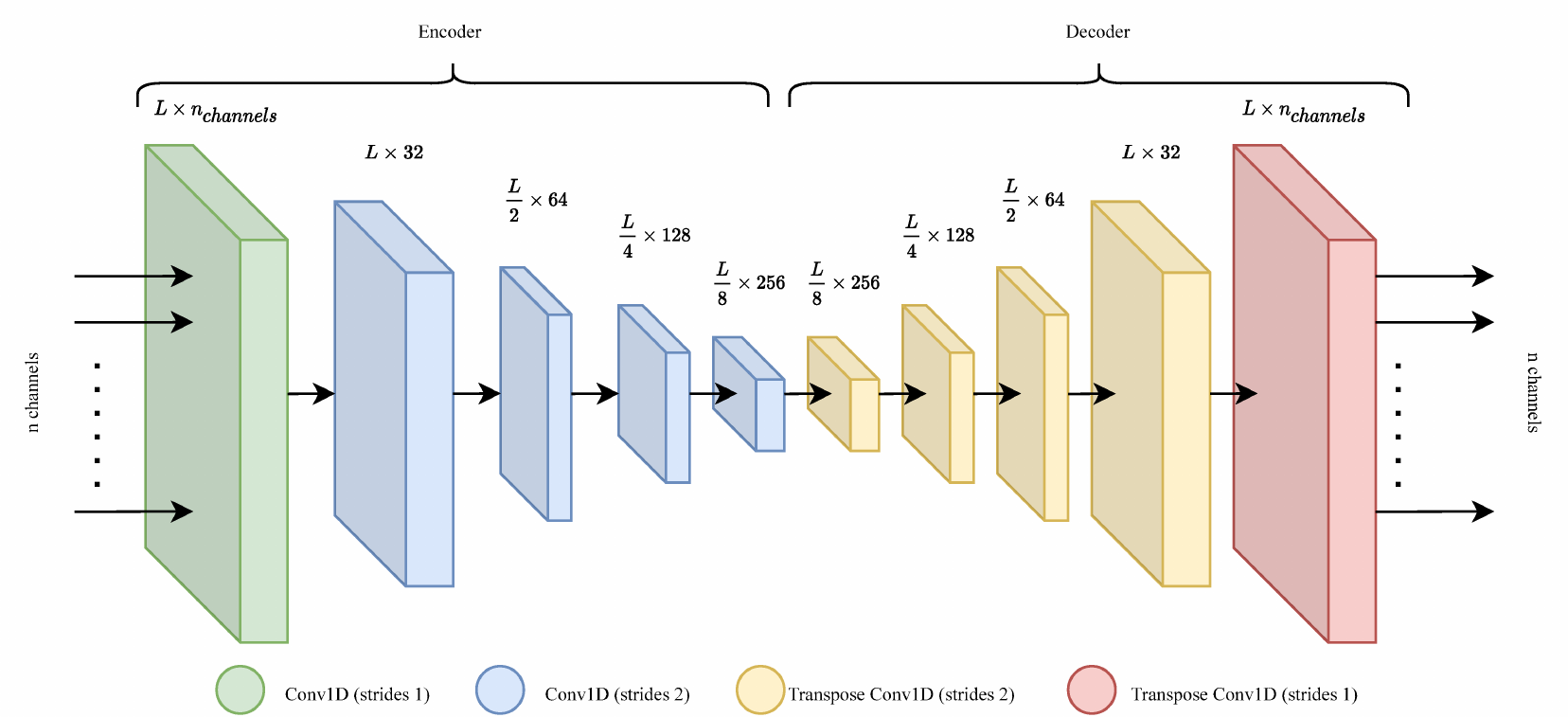}
   }}
  \caption{The Fully Convolutional Denoising Autoencoder architecture.}
  \label{figure:FC-DAE_architecture}
\end{figure*}

An end-to-end Fully Convolutional Denoising Autoencoder was used in order to filter the noisy input signal. The FC-DAE receives as input the multichannel recording and produces a clean multichannel output containing only the neuronal activity signal. The network architecture is illustrated in Figure \ref{figure:FC-DAE_architecture}. In this work, we completely omit the Fully Connected layer that is found in CNNs, adopting the rationale proposed in \cite{chiang2019noise}. This approach presents important benefits, mainly being computationally efficient and preserving the local information \cite{fu2017raw}. Additionally, we do not include any MaxPooling layers, since they may fail to preserve important structural details \cite{chiang2019noise}. The network's encoder composes of five one-dimensional convolutional layers while the decoder has a symmetric structure composed of five one-dimensional transposed convolutions. 

The input recording is initially zero-meaned and then normalized in the $[-1,1]$ range. Since the FC-DAE operates on batches of data, the recording is split into samples using non-overlapped windows of size L. The FC-DAE jointly processes the channels, taking advantage of their interrelations, and outputs the n-channel filtered sample of length L.

The window length was selected to be 512 samples, which corresponds to a duration of 51.2ms considering the frequency sampling of $10kHz$. The number of channels depends on the probe used for recording the extracellular action potentials (EAP).

The kernel size was set to 3 for all convolutional layers, since after experimentation it was observed that the small kernel achieved good results and ensured computational efficiency thanks to the small number of parameters. The ELU activation function was used for all the layers with the exception of the last one where the linear activation was used. The learning rate was set to 0.001 and the Mean Square Error loss function was utilized.

\subsection{SEAset: Simulated Extracellular Action Potential Dataset}
For the needs of this study a simulated dataset of extracellular recordings was generated. Since the supervised learning scheme is employed to train the FC-DAE, a ground truth clean recording is necessary. The patch-clamp approach has been used as a method to provide real-life EAPs together with the corresponding intracellular signal that can be considered as the ground truth. However, they are limited by the small number of cells that can be recorded simultaneously \cite{buccino2021mearec}. Thus, we opted for the simulation of the recordings to ensure control over the recording conditions. 

The EAPs were simulated with the MEArec Python library \cite{buccino2021mearec}. Neuronal models of the L5 area were used to simulate the activity of three inhibitory and three excitatory neurons. To ensure the generation of realistic recordings the distance-correlated noise was added as described in \cite{buccino2021mearec} and the bursting and overlap settings were activated.

A simulated tetrode probe was selected to record the EAPs resulting in a four-channel input signal. The sampling frequency was set to $10 kHz$. The ground truth clean signal was generated by aggregating the neuronal activity of all six cells into one single signal. Since the network's output is multichannel, the same aggregated signal was used as ground truth for all four channels.

To evaluate the robustness of our approach under different levels of noise, we generated signals contaminated with noise standard deviation (\cite{buccino2021mearec}) of 7, 9, 15 and 20 $\mu V$. For each noise level a set of ten 5s recordings was generated and for every recording, a randomly generated set of seeds was used for the MEArec's spiketrains, templates, convolution and noise seeds. For a better intuition on the SEAset recordings' noise contamination, we examined the observed SNR values. Let $\hat{X}_{i_{rec}, l}^n$ be the $l-th$ noisy channel of the $i_{rec}-th$ recording contaminated with noise of level $n\ \mu V$, where $l \in [1, 2, 3, 4]$, $i_{rec} \in [1, ..., 10]$  and $n \in [7, 9, 15, 20]\ \mu V$. The corresponding ground truth clean signal is noted as $X_{i_{rec}, l}^n$. We calculate the observed SNR of $\hat{X}_{i_{rec}, l}^n$ as:
$$
SNR_{i_{rec}, l}^n = 10\log_{10}{\frac{\sum(X_{i_{rec}, l}^n)^2}{\sum(\hat{X}_{i_{rec}, l}^n - X_{i_{rec}, l}^n)^2}}
$$
We then calculate the average $SNR_{l_{avg}}^n$ across all ten recordings for each channel $l$, 
$
SNR_{l_{avg}}^n = \frac{1}{10}\sum_{i_{rec}}{SNR_{i_{rec}, l}^n}
$.
Finally, for each noise level $n\ \mu V$ we calculate the minimum and maximum SNR values across all four channels (Table \ref{table:seaset_snr}). 

\begin{table}[h]
\caption{Mimimum and Maximum observed SNR for each noise level}
\label{table_example}
\begin{center}
\begin{tabular}{|c||c|c|}
\hline
Noise Level ($\mu V$) & Minimum SNR ($dB$) & Maximum SNR ($dB$)\\
\hline
7  &  0.60 & 2.87  \\
9  & -1.36 & 0.77  \\
15 & -5.29 & -3.59 \\
20 & -7.02 & -5.97 \\
\hline
\end{tabular}
\end{center}
\label{table:seaset_snr}
\end{table}

\subsection{DWT and SWT filtering methods}
Our method was evaluated in comparison with two wavelet denoising methods, the Discrete Wavelet Transform (DWT) and the Stationary Wavelet Transform (SWT). In general, all wavelet-based denoising techniques involve three steps: 1)~Decomposition: Choosing a mother wavelet and a maximum decomposition level and then decomposing the signal to obtain the detail and approximation coefficients for each level, 2)~Thresholding: Computing threshold values and applying soft or hard threshold to the detail coefficients for each level, 3)~Reconstruction: Reconstructing the signal by using the modified coefficients. 

The main difference between DWT and SWT lies upon the decimation operation. Specifically, the number of coefficients of each level is half that of the preceding level in DWT, while the number of coefficients is the same for each level in SWT. The retention of redundant data renders the SWT translation-invariant, which is a useful property for denoising neural recordings~\cite{Farina2013}.

The effectiveness of wavelet denoising algorithms strongly depends on the implementation choices at the different steps, including the decomposition level, the mother wavelet, the threshold method and its type (i.e., how it is applied). For this reason, we explored several combinations of these parameters during the parameter tuning process. Inspired by \cite{Baldazzi2020}, we explored five orthogonal and biorthogonal mother wavelets: Haar, Coiflet2 (Coif2), Daubechies4 (Db4), Biorthogonal6.8 (Bior6.8) and Symlet7 (Sym7). Regarding the threshold selection method, we explored four options~\cite{Ergen2012, Johnstone1997}: universal, minimax, SURE and heuristic SURE. The aforementioned threshold selection methods were combined with soft and hard type of thresholding. Furthermore, the signals were decomposed at four and five decomposition levels. Consequently, taking into account the possible values of the four implementation choices, our parameter space consisted of 80 different combinations.

\section{Experimental Setup}
We train and evaluate our model on the SEAset data employing the leave-one-out (LOO) strategy. This enables the evaluation of the model on previously unseen recordings and the examination of its generalization ability. For each noise level, the evaluation process was repeated ten times; each time, the training set consisted of nine 5s recordings, which provided a sufficient amount of samples for training, and the test set consisted of one 5s recording. 

The same LOO evaluation process was followed for the parameter selection of the DWT and SWT wavelet denoising methods. Specifically, in each LOO iteration, the parameters were tuned on the training set by means of searching the parameter combination which minimizes the Mean Square Error between the denoised and the ground truth signal. Then, the performance of the methods was evaluated on the test set. 

The per channel SNR improvement \cite{fotiadou2020multi} and Root Mean Square Error (RMSE) are utilized as performance metrics. Each value occurs as the average of all ten iterations of the LOO process respectively. The SNR improvement provides an intuition on how well the filtering method removed the input signal's noise, while the RMSE evaluates the similarity between the method's output and the ground truth. This is important in spike sorting applications where the original action potential form can contain useful information about the source neuron and should be maintained by the preprocessing filtering method.

\section{Experimental Results}
\begin{figure*}[htbp]
  \centering
  \framebox{\parbox{6.5in}{
  \centering
  \includegraphics[scale=0.25]{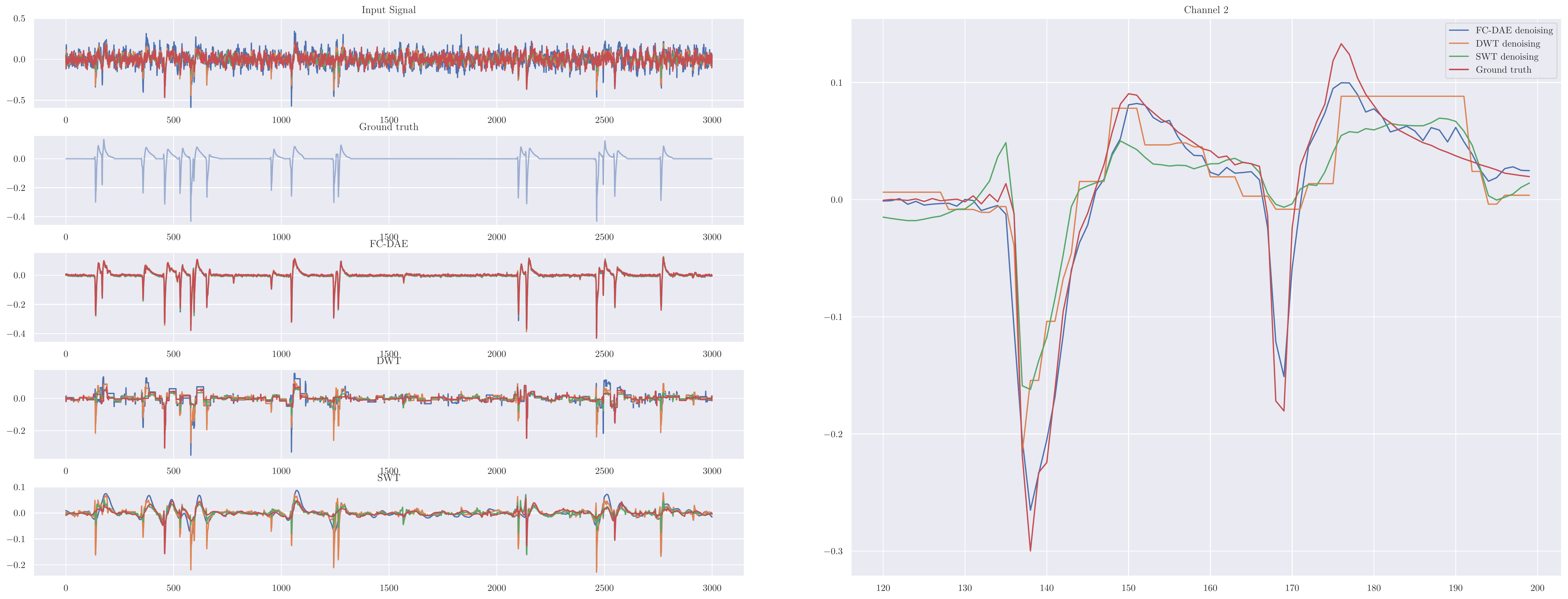}
   }}
  \caption{Left: Example output of the FC-DAE, DWT and SWT methods for a section of recordings of noise level $9\ \mu V$. Right: Recording details from sample 120 to 200.}
  \label{figure:example_output}
\end{figure*}

Figure \ref{figure:example_output} presents an example of the denoising outputs for the FC-DAE, DWT and SWT and allows a qualitative examination of their performance. In the left column, a section of a test recording of noise $9\ \mu V$ is processed by the three methods. The top plot displays the noisy input, while the second one the ground truth signal. The following three plots present the four-channel outputs of the FC-DAE, the DWT and the SWT respectively. The DWT and SWT output signals are pretty similar and manage to isolate the spikes relatively well. The FC-DAE's output, however, is visibly cleaner. In areas of no spiking activity, the autoencoder manages to completely suppress the noise, in contrast to the wavelet denoising methods where some noise is still present. At the same time, the FC-DAE's spikes' shape is cleaner and quite close to the ground truth.

A close-up look at the signals' section between the samples 120 and 200 is plotted in the right column of Figure \ref{figure:example_output}. Here, two consecutive EAPs have been recorded, with very little space between them. For legibility reasons, we only display the channel in which the DWT and SWT better performed, i.e. the second one. The FC-DAE output presents significantly more similarity to the ground truth, compared to the DWT and SWT signals, and did not fail to produce the second spike, unlike the wavelet methods.

\begin{figure}[htbp]
  \centering
  \framebox{\parbox{3in}{
  \centering
  \includegraphics[scale=0.42]{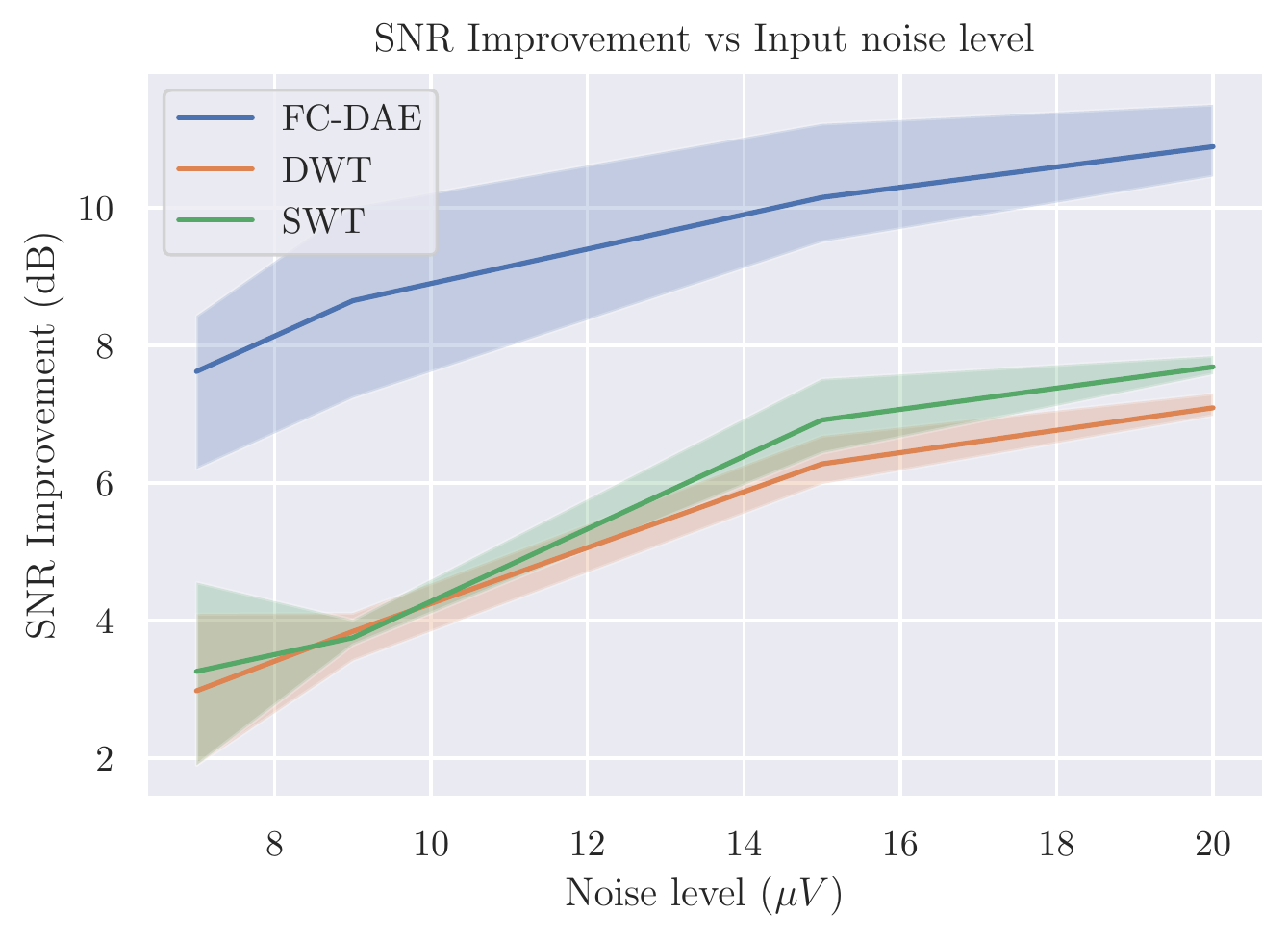}
  \includegraphics[scale=0.42]{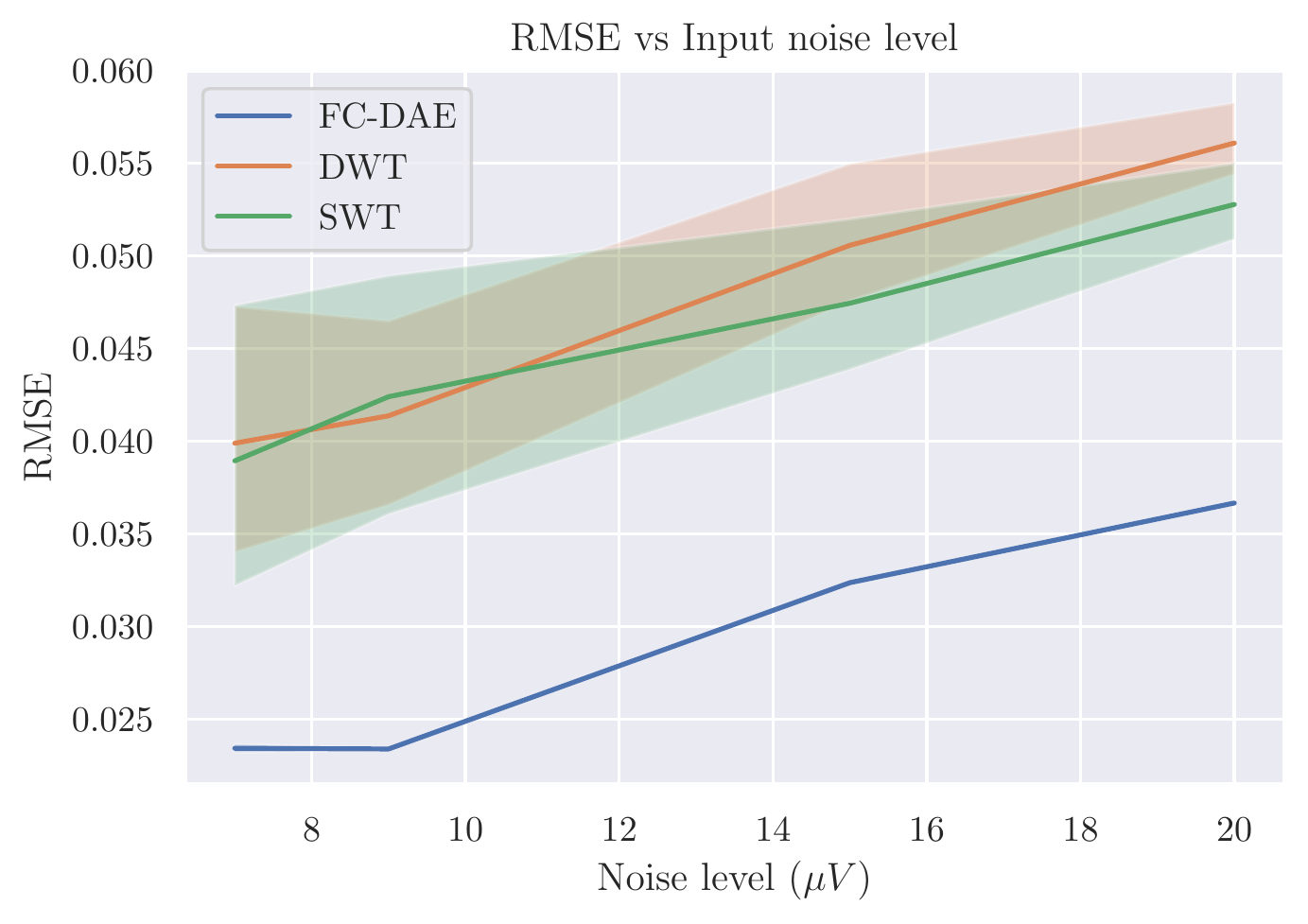}
   }}
  \caption{Up: Average SNR Improvement across all channels for each method (bold line) and the minimum and maximum values (transparent range). Down: Similarly for each method's RMSE performance.}
  \label{figure:per_channel_result_summary}
\end{figure}

The SNR improvement and RMSE for all four examined noise levels are summarized in Figure \ref{figure:per_channel_result_summary}. Here, the initial comments made above are confirmed, for all noise levels. The FC-DAE manages to outperform the wavelet denoising methods both in terms of denoising and reconstructing the clean neuronal activity signal. Additionally, the FC-DAE presents significantly less variance in its estimations' RMSE across the four channels, in contrast to the DWT and SWT.

\section{Conclusion}
In this work, a Deep Fully Convolutional Denoising Autoencoder was developed for denoising EAP recordings. A realistic simulated dataset of neuronal recordings from the L5 area was created and then was used to train and evaluate the network. In the experiments conducted, our proposed method showed that it can achieve a substantial quality improvement of noisy recordings with various noise levels while simultaneously contributing to the effective reconstruction of the EAP. The proposed FC-DAE presented superior performance in comparison with the widely-used wavelet denoising methods of DWT and SWT. Future work will include denoising real recordings using the FC-DAE trained on simulated datasets. Further, evaluating the performance of spike sorting on FC-DAE pre-processed recordings will allow a more thorough examination of the denoising robustness. Finally, it should be investigated whether one model can handle EAPs from different brain areas with different spike shapes, or an area-specific model approach would be more appropriate.  

\addtolength{\textheight}{-12cm}   



\bibliographystyle{IEEEtran}
\bibliography{IEEEabrv,reference_list.bib}

\end{document}